\pdfoutput=1

\documentclass[aps,twocolumn,showpacs,amssymb,amsmath,nofootinbib,floatfix,superscriptaddress,showkeys]{revtex4-1}

\usepackage{graphicx,color}
\usepackage{amsmath,amssymb,bm}
\usepackage{float}
\usepackage{latexsym}
\usepackage{comment}
\usepackage{hyperref}
\usepackage{epsfig}

\begin{document}

\title{Proton-proton hollowness at the LHC from inverse scattering}

\author{Enrique Ruiz Arriola}
\email{earriola@ugr.es}
\affiliation{Departamento de F\'isica At\'omica, Molecular y Nuclear and Instituto Carlos I de
  Fisica Te\'orica y Computacional,  Universidad de Granada, E-18071
  Granada, Spain}

\author{Wojciech Broniowski}
\email{Wojciech.Broniowski@ifj.edu.pl}
\affiliation{The H. Niewodnicza\'nski Institute of Nuclear Physics, Polish Academy of Sciences, 31-342 Cracow, Poland}
\affiliation{Institute of Physics, Jan Kochanowski University, 25-406 Kielce, Poland}

\begin{abstract}

Parameterizations of the pp scattering data at the LHC collision
energies indicate a hollow in the inelasticity profile of the pp
interaction, with less absorption for head-on collisions than at a
non-zero impact parameter.  We show that some qualitatively unnoticed
features may be unveiled by a judicious application of the inverse
scattering problem in the eikonal approximation and interpretation
within an optical potential model. The hollowness effect is magnified
in a 3D picture of the optical potential, and will presumably be
enhanced at yet higher energies. Moreover, in 3D it sets in at much
smaller energies than at the LHC.  We argue that hollowness in the
impact parameter is a quantum effect, relying on the build-up of the
real part of the eikonal scattering phase and its possible passage
through $\pi/2$. We also show that it precludes models of inelastic
collisions where inelasticity is obtained by naive folding of partonic
densities.
\end{abstract}

\date{ver. 3, 30 January 2017}

\pacs{13.75.Cs, 13.85.Hd}

\keywords{proton-proton collisions, elastic and inelastic cross section, inverse scattering, Glauber model}

\maketitle

\section{Introduction \label{sec:intro}}

The main purpose of scattering experiments is to unveil the underlying
structure of the colliding particles. However, there is always a
limiting resolution of the relative de Broglie wavelength 
\mbox{$\Delta r = 1/p_{\rm CM} \sim 2/\sqrt{s}$}, which effectively coarse-grains both the
interaction between colliding particles and their structure as seen in
the collision process. Of course, as the energy increases, new
production channels open and inelasticities become important, but
this does not change the overall picture even if the elastic
scattering is regarded as the diffractive shadow of the particle
production. Besides the early cosmic rays investigations in the mid
50's~\cite{williams1955highII} reporting a surprisingly {\it too large
cross section} compared to accelerator
extrapolations~\cite{williams1955highI}, the accumulation of more
precise scattering data since the early 60's until the ISR experiments
in the 70's (see, e.g.,~\cite{Amaldi:1979kd} for a compilation), has
been modifying our picture of the nucleon along the years%
~\cite{Block:1984ru,Cheng:1987ga,Matthiae:1994uw,Barone:2002cv,Dremin:2012ke}
with the deceiving result that the asymptotic regime may still be
further away than hitherto assumed.  The shortest wavelengths ever
available in a terrestrial laboratory are achieved in the current and
upcoming proton-proton (pp) scattering at the CERN Large Hadron
Collider, with \mbox{$\sqrt{s}= 7-14 {\rm ~TeV}$} corresponding to
$\Delta r \sim 0.001 {\rm ~fm} = 1 {\rm ~am}$, a tiny length compared
to the conventional proton size. From the point of view of the
relative distance, the maximum momentum transfer $t_{}=-\vec
q_\perp^2$ samples the smallest impact parameter $\Delta
b=1/q_{\perp}$. The succinct summary of the whole development is that,
historically, protons become larger, edgier and blacker as the energy
of the collision is being increased.

In a recent communication~\cite{Arriola:2016bxa}, we have analyzed the
TOTEM data~\cite{Antchev:2013gaa} for the pp collisions at
$\sqrt{s}=7$~TeV in terms of the so-called {\em on-shell optical
potential}. A striking result is that there appears to be more
inelasticity when the two protons are at about half a fermi traverse
separation than for head-on collisions: a hollow is developed in the
pp inelasticity profile. This counterintuitive finding has also been
noticed by several other
authors~\cite{Alkin:2014rfa,Dremin:2014eva,Dremin:2014spa,Anisovich:2014wha,Dremin:2016ugi}. As
we will show, it actually precludes a probabilistic geometric
explanation of the pp inelasticity profile based on folding of
one-body partonic densities. We note that microscopic realization of
the hollowness effect has been offered within a hot spot Glauber
model~\cite{Albacete:2016pmp} for the elastic pp amplitude.

In the present paper we largely extend the findings of
Ref.~\cite{Arriola:2016bxa}.  We analyze the problem from an
inverse-scattering point of view, utilizing the standard {\em optical
potential} in the eikonal approximation (not to be confused with the
on-shell one, see below). The eikonal method is justified for
sufficiently small impact parameters, $b < 3 {\rm ~fm}$, and for the
CM energies of the system $\sqrt{s}> 20 {\rm ~GeV}$.  The 3D hole in
the optical potential emerges already at $\sqrt{s} \sim 1 {\rm ~TeV}$,
well below the present LHC energies. We note that the hollowness
effect becomes less visible in the 2D inelasticity profile in the
impact parameter space, where geometrically the 3D hole is covered up
by the accumulated longitudinal opacity of two colliding protons.

We take no position on the particular underlying dynamics of the
system. Instead, we rely on
accepted and working parameterizations of the NN scattering amplitude.
For definiteness, we apply the modified Barger-Phillips amplitude~2
(MBP2) used in the comprehensive analysis of Fagundes {\it et
  al.}~\cite{Fagundes:2013aja}, where the implemented properties at
low- and high values of $t$ are indeed supported by reasonable
$\chi^2$ values and visual inspection vs data. It is thus fair to
assume that these fits capture the essence of the scattering amplitude
at any fixed energy and up to a certain $t_{\rm max}$.
Correspondingly, the present experimental range covers impact
parameters larger than $b_{\rm min} \sim 0.1 {\rm ~fm}$, which is the
fiducial domain of the present study.

We use the well-established inverse scattering methods to determine
the optical potential. This has the advantage of being free of
dynamical assumptions, in particular, naive folding features assumed
quite naturally by model calculations but which turn out to be hard to
reconcile with the hollowness effect.

Finally, let us note that we will not make any separation other than
single elastic channel from inelastic channels (being all the
rest). Therefore the verification of the conjecture that the
calculated elastic cross section includes diffraction, whereas the
inelastic cross section only includes uncorrelated processes, as put
forward in Refs.~\cite{Lipari:2009rm,Lipari:2013kta,Fagundes:2015vba},
will not be addressed in the present study.

\section{Mass squared approach with central optical potential}

The NN elastic scattering amplitude has 5 complex Wolfenstein
components, as it corresponds to scattering of identical spin 1/2 particles~\cite{PhysRev.85.947}. 
Besides, at high energies, \mbox{$\sqrt{s} \gg 2 M_N$},  both relativistic
effects and inelasticities must be taken into account. In principle, a field
theoretic description of particle production would require solving a
multi-channel Bethe-Salpeter (BS) equation. Taking into account that most of
the produced particles are pions, the maximum number of coupled
channels involving just direct pion production $pp \to pp + n \pi$ 
necessary to preserve the (coupled channel) unitarity would involve at least
$n^{\rm max} \sim (\sqrt{s}-2 M_N)/m_\pi$ channels.  For ISR energies it
corresponds to $n_{\rm ISR}^{\rm max} \sim 150-450$, whereas for the
LHC energies $n_{\rm LHC}^{\rm max} \sim 5 \times 10^5$. Such a huge
number of channels prevents from the outset a direct coupled channel
calculation.\footnote{Of course, the average number of produced particles 
is estimated to be much smaller, $N=\langle n (\sqrt{s})
\rangle \sim 0.88 + 0.44 \log (s/s_0) + 0.118 \log^2 (s/s_0)$
($\sqrt{s_0} = 1 {\rm
~GeV}$)~\cite{Thome:1977ky,GrosseOetringhaus:2009kz}, which becomes
$\sim 8-12$ for ISR and $\sim 50$ for the LHC, but one does not know
how to pick the relevant ``averaged'' combinations of coupled channels
to apply the BS method.}  Another added difficulty is the
incorporation of spin at these high energies, mainly because the
experimental information is insufficient. Thus, as it is usually
assumed in most calculations, at these high energies spin effects are
fully neglected and a purely central type of interaction is taken.

An advantageous way to take into account inelasticities is to recourse
to an optical potential where all inelastic channels are in principle
integrated out. Even if all the particle production
processes were known, an explicit construction for the huge number of
channels has never been carried out, hence our approach is
phenomenological, with the idea to deduce the optical potential 
directly from the data via an inverse scattering method. 
Because such a framework is currently not commonly used in 
high-energy physics, it is appropriate to
review it here, providing in passing a justification on why we choose it. 

The optical potential was first introduced to describe the inelastic
neutron-nucleus scattering above the compound nucleus
regime~\cite{Fernbach:1949zz} (typically in the $10-500 {\rm ~MeV}$
range).  There, the concept of the black disk limit was first tested,
along with the Fraunhofer diffraction pattern appearing as a shadow
scattering effect. This work inspired Glauber's seminal
studies~\cite{glauber1959high} on the eikonal approximation, which is
currently successfully applied to model the early stages of the
ultra-relativistic heavy-ion collisions (see
e.g. \cite{Florkowski:2010zz}).
Serber~\cite{serber1963theory,serber1964scaling,serber1964high}
provided an extension of the optical eikonal formalism to high energy
particle physics.  As it was shown by Omnes~\cite{omnes1965optical},
the simple assumption of a double spectral representation of the
Mandelstam representation of the scattering amplitude suffices to
justify the use of an optical potential. Cornwall and
Ruderman~\cite{cornwall1962mandelstam} delineated a more precise
definition of the optical potential, directly based in field theory
and tracing its analytic properties from the causality
requirement. Some further field theoretic discussions using the
multichannel BS equation can be found
in~\cite{torgerson1966field,arnold1967optical}, and were early
reviewed by Islam~\cite{islam1972optical}.

The simplest way of retaining relativity without solving a BS equation
with a {\it phenomenological} optical potential is by using the
so-called mass squared method, discussed by Allen, Payne, and Polyzou
in an insightful paper~\cite{Allen:2000xy}.\footnote{These authors
proposed a practical way to promote non-relativistic fits of NN
scattering to a relativistic formulation without a necessity of refitting
parameters.}
The idea is to postulate the total squared mass operator for the pp system as
\begin{eqnarray}
{\cal M}^2 = P^\mu P_\mu  \stackrel{CM}{=}4(p^2+M_N^2) + {\cal V} ,
\end{eqnarray}
where  $P^\mu$ is the total four-momentum, CM indicates the center-of-mass frame, 
$p$ is the CM momentum of each nucleon, $M_N$ is the nucleon mass, and
${\cal V}$ represents the invariant (momentum-independent) interaction, whose form can be
determined in the CM frame by matching to the non-relativistic limit
with a non-relativistic potential $V(\vec x)$. This allows one,
after quantization,  to write down the relativistic wave equation 
${\cal \hat M}^2 \Psi = s \Psi $, 
in the form of an equivalent non-relativistic Schr\"odinger equation~\cite{Allen:2000xy}
\begin{eqnarray}
(-\nabla^2 +U ) \Psi = (s/4-M_N^2) \Psi ,
\label{eq:mass-sq}
\end{eqnarray}
with the reduced potential $U= M_N V$.
In essence, the invoked prescription corresponds to a simple rule where one may effectively implement
relativity by just promoting the non-relativistic CM momentum to the relativistic CM momentum.

\begin{figure*}
\begin{center}
\includegraphics[width=0.35\textwidth]{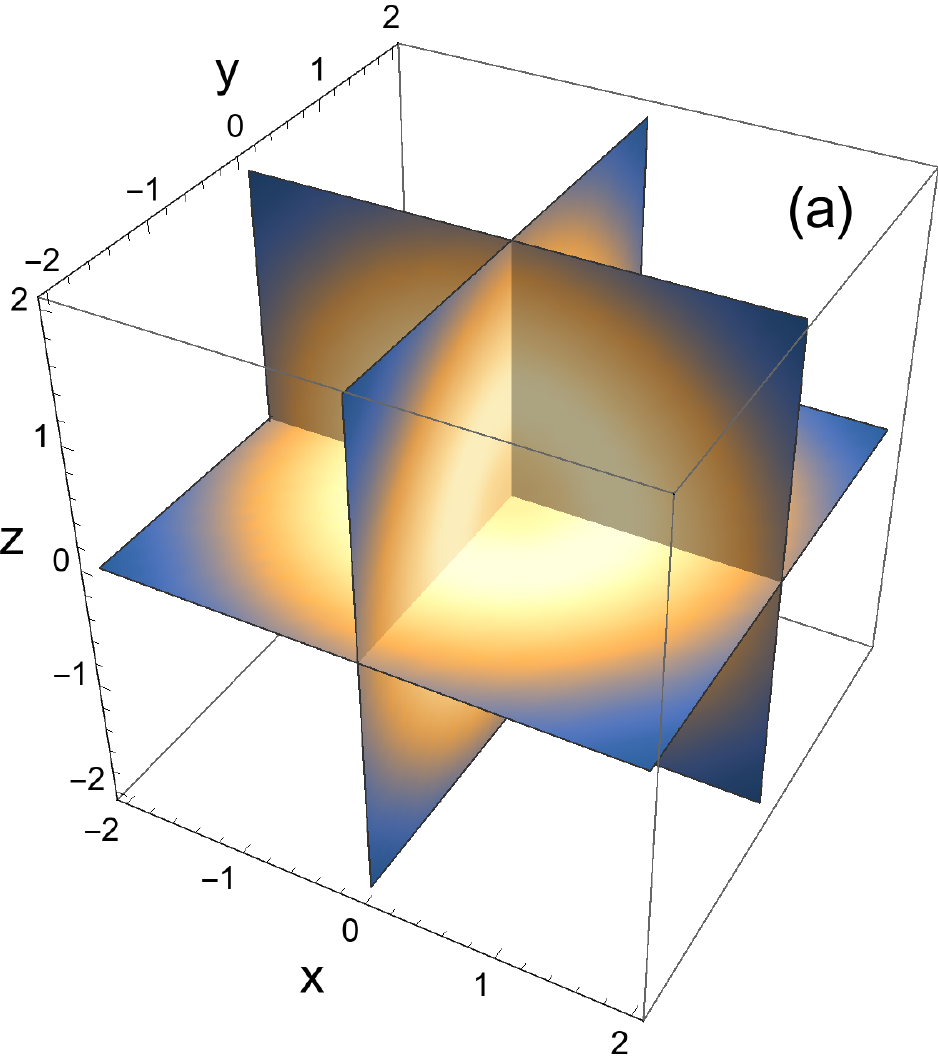}\hspace{1cm}  \includegraphics[width=0.35\textwidth]{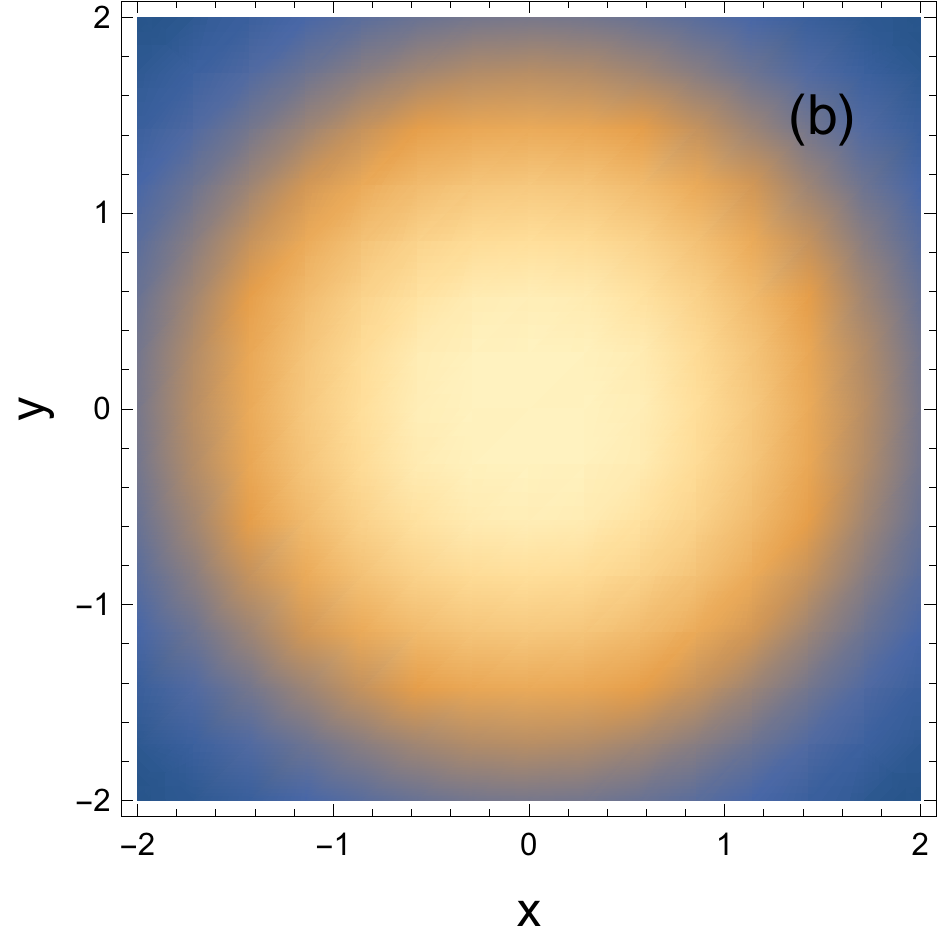}
\end{center}
\vspace{-5mm}
\caption{Projection of a sample spherically symmetric three-dimensional function (a)  on two dimensions (b), as in Eq.~(\ref{eq:nin1}).  A shallow 
hollow present in (a) disappears in (b), where it is only reflected with a flatness on the central region. \label{fig:3D2D}}
\end{figure*}

As remarked by Omnes~\cite{omnes1965optical},
``one can always find an optical potential that fits any amplitude
satisfying the Mandelstam analyticity assumptions'', and we apply a definite prescription to accomplish this goal.
To account for inelasticity, 
we assume an energy-dependent and local phenomenological
optical potential, \mbox{$U(\vec r;s)= {\rm Re }\, U(\vec r;s) + i {\rm Im}\, U(\vec r;s)$}, which can be obtained by fitting the scattering
data. Due to causality, the optical potential in the $s$ channel
satisfies a fixed-$r$ dispersion relation. Together with Eq.~(\ref{eq:mass-sq}), it provides the
necessary physical ingredients present in any field
theoretic approach: relativity and inelasticity, consistent with
analyticity. The potential $U$ appearing in Eq.~(\ref{eq:mass-sq}) will be
determined in the following via inverse scattering in the eikonal
approximation for any value of $s$. To ease the notation,
the $s$-dependence is suppressed below. 

\section{On-shell optical potential and the eikonal approximation}

Besides the ``standard'' potential $U$, the object we are going to use
is the {\em on-shell optical potential} $W$, defined by a Low-type
integral equation discussed, e.g.,
in~\cite{cornwall1962mandelstam,namyslowski1967relativistic,Nieves:1999bx,Arriola:2016bxa}.
From Eq.~(\ref{eq:mass-sq}) we get for the probability flux 
\begin{eqnarray} 
\oint_{r=R} \vec dS \cdot \vec J  = \int_{r \le R} d^3 x \, {\rm Im}\, U (\vec x) |\Psi(\vec x)|^2,  
\end{eqnarray} 
with $\vec J= \Psi^* (\vec \nabla \Psi) - (\vec \nabla \Psi^*) \Psi$
denoting the probability current.  The asymptotic behavior of the wave
function is $\Psi(\vec x) \to e^{i \vec p \cdot \vec x} + f (\hat x)
{e^{i p r}}/{r}$.  It follows from the definition of the inelastic
cross section that
\begin{eqnarray}
\sigma_T - \sigma_{\rm el} \equiv \sigma_{\rm in} = - \frac{1}{p} \int d^3 x \, {\rm Im} \, U( \vec x ) |\Psi(\vec x) |^2,
\end{eqnarray} 
with shows that the density of inelasticity is proportional to the
absorptive part of the optical potential times the square of the
modulus of the wave function.  One can now identify the on-shell
optical potential\footnote{An interesting observation of Cornwall and
  Ruderman~\cite{cornwall1962mandelstam} was that the on-shell optical
  potential does not involve the wave function itself.} as
\begin{eqnarray}
{\rm Im} \, W( \vec x ) = {\rm Im} \, U( \vec x ) |\Psi(\vec x) |^2 .
\end{eqnarray} 
In the eikonal approximation one has
\begin{eqnarray}
\Psi (\vec x)= \exp\left[ i p z - \frac{i}{2p}\int_{-\infty}^z U(\vec b,z') dz' \right],
\end{eqnarray} 
thus 
\begin{eqnarray}
{\rm Im} \, W( \vec x)  = p \frac{d}{dz} \exp\left[ \frac{1}{p}\int_{-\infty}^z {\rm Im} U(\vec b,z') dz'\right]. \label{eq:WU}
\end{eqnarray} 
Upon $z$ integration,
\begin{eqnarray}
-\frac{1}{p}\int_{-\infty}^\infty dz {\rm Im} \, W( \vec b, z) =  1- e^{- 2 {\rm Im} \chi (b)} \equiv n_{\rm in}(b), \label{eq:nin1}
\end{eqnarray} 
where 
\begin{eqnarray}
\hspace{-5mm} \chi (b)=- \frac{1}{2p} \int_{-\infty}^\infty U(\sqrt{b^2+z^2}) dz = - \frac{1}{p} \int_{b}^\infty \frac{r  U(r)\,dr}{\sqrt{r^2-b^2}} \nonumber \\ \label{eq:chi}
\end{eqnarray}
is the (complex) eikonal phase~\cite{glauber1959high}.
Equation~(\ref{eq:WU}) is the standard result for the inelasticity
profile $n_{\rm in} (b)$ in the eikonal approximation.\footnote{Alternative
  eikonal unitarization schemes to the standard one have been
  suggested long ago~\cite{Blankenbecler:1962ez}, but they do not
  fulfill the above relation.} Note that it links the imaginary part
of the eikonal phase with the absorptive part of the on-shell optical
potential $W$, hence the significance of this object in the present study.

The inverse scattering problem has been solved
in~\cite{newton1962construction} and in the eikonal approximation
in~\cite{omnes1965optical} (for a review see,
e.g.,~\cite{buck1974inversion}). In our case the inversion is based on
the fact that Eq.~(\ref{eq:chi}) is of a type of the Abel integral
equation, hence the solution for the optical potential $U$ takes the
simple form~\cite{glauber1959high}
\begin{eqnarray}
U(r) = M_N V(r)= \frac{2p}{\pi} \int_r^\infty db \frac{\chi'(b)}{\sqrt{b^2-r^2}}, \label{eq:Uchi}
\end{eqnarray}
which may be straightforwardly checked via direct
  substitution~\footnote{We use a slightly different form than the
  original Glauber formulation~\cite{glauber1959high}, more suitable
  for numerical work, since care must be exercised with the handling of
  derivatives at the end-point singularity at $b=r$. Our form was used in the
  NN analysis of Ref.~\cite{Arriola:2014lxa}.}. Similarly, from
  Eq.~(\ref{eq:nin1}) one obtains
\begin{eqnarray}
{\rm Im} W(r) = \frac{p}{\pi} \int_r^\infty db \frac{n_{\rm
in}'(b)}{\sqrt{b^2-r^2}}. \label{eq:Wn}
\end{eqnarray}
As the (complex) scattering phase may be obtained
from the data parameterizations (see the following section), Eqs.~(\ref{eq:Uchi}) and (\ref{eq:Wn}) provide a simple way to obtain the corresponding optical potentials.
An investigation of their behavior with the increasing collision energy is our principal goal.

Before going to the details of the next sections, let us comment on a
simple geometric interpretation of formula~(\ref{eq:nin1}).  Suppose
we have a spherically symmetric three-dimensional function with a
lower density in the middle than in outer layers, as depicted in
Fig.~\ref{fig:3D2D}(a). If the hollow is not too deep, the projection
of the function on two dimensions, as presented in Eq.~(\ref{eq:nin1}), 
covers it up by the inclusion of the outer layers. 
In the example of Fig.~\ref{fig:3D2D}(b) the central region
is flat. Therefore the flatness of the inelasticity profile
$n_{\rm in}(b)$ corresponds to a hollow in the imaginary part of the on-shell
optical potential ${\rm Im}W(r)$. In other words, the three-dimensional
objects as ${\rm Im}W(r)$ or $U(r)$ are more sensitive to exhibit a
hollow than their corresponding 2D projections, i.e., the inelasticity profile or the eikonal
phase.

\section{Amplitudes and parameterization \label{sec:param}}

The pp elastic scattering differential cross section is given by 
\begin{eqnarray}
\frac{d\sigma_{\rm el}}{dt}= \frac{\pi}{p^2} \frac{d \sigma_{\rm el}}{d \Omega} = 
\frac{\pi}{p^2} |f(s,t) |^2 \,  , 
\end{eqnarray}
with the spinless partial wave expansion of the scattering amplitude 
\begin{eqnarray}
&&f(s,t) =\sum_{l=0}^ \infty (2l+1) f_l(p) P_l(\cos \theta)  \label{eq:PWA} \\
&&=  \frac{p^2}{\pi} \int d^2 b \, h(\vec b,s) \, e^{i \vec q \cdot \vec b} = 
 2 p^2 \int_0^\infty b db J_0(bq) h(b,s) \, ,  \nonumber
\end{eqnarray} 
where $t=-\vec q^2$ and  \mbox{$q = 2p \sin
(\theta/2) $} denotes the momentum transfer. The Coulomb effects can be neglected for $|t| > 8 \pi
\alpha /\sigma_T$ ($\alpha \simeq 1/137.04$ is the fine structure
constant)~\cite{Block:1984ru}.  In the eikonal limit, justified for $
p a \gg 1$ with $a$ standing for the range of the interaction, one has $bp = l +1/2
+ {\cal O}(s^{-1})$, hence the amplitude in the impact-parameter
representation becomes
\begin{eqnarray}
h(b,s)=\frac{i}{2p} \left [ 1-e^{i \chi(b)}  \right ] =f_l(p) + {\cal O}(s^{-1}), \label{eq:eik}
\end{eqnarray}
whereas $P_l(\cos \theta) \to J_0 (qb)$.
Explicitly, 
\begin{eqnarray}
2p h(s,b) = \frac{1}{p} \int_0^\infty q dq J_0(bq) f(s,-q^2). \label{eq:invf}
\end{eqnarray} 
The standard formulas for the total, elastic, and total inelastic cross sections read~\cite{Blankenbecler:1962ez}
\begin{eqnarray} 
\sigma_T  &=& \frac{4 \pi}p {\rm Im} f(s,0) = 4 p \int d^2 b {\rm Im} h(\vec b,s) \nonumber \\ 
                &=& 2 \int d^2 b \left [ 1- {\rm Re} \, e^{i \chi(b)} \right ] \, \label{eq:st} ,\\
\sigma_{\rm el} &=& \int d\Omega |f(s,t)|^2 = 4 p^2 \int d^2 b |h(\vec b,s)|^2 \nonumber \\ 
                &=&  \int d^2 b | 1- e^{i \chi(b)} |^2 \, \label{eq:sel}, \\
\sigma_{\rm in} &\equiv& \sigma_T - \sigma_{\rm el} = \int d^2 b n_{\rm in} (b) \nonumber \\ 
                &=&  \int d^2 b \left [ 1- e^{- 2 {\rm Im} \chi (b)} \right ] \, . \label{eq:sin}
\end{eqnarray} 
The inelasticity profile 
\begin{eqnarray}
n_{\rm in} (b)  = 4p {\rm Im} h(b,s) -  4p^2|h(b,s)|^2,   \label{eq:prof}
\end{eqnarray} 
satisfies $n_{\rm in}(b) \le 1$, conforming to unitarity and the probabilistic interpretation of absorption.
\begin{figure}
\begin{center}
\includegraphics[width=0.35\textwidth]{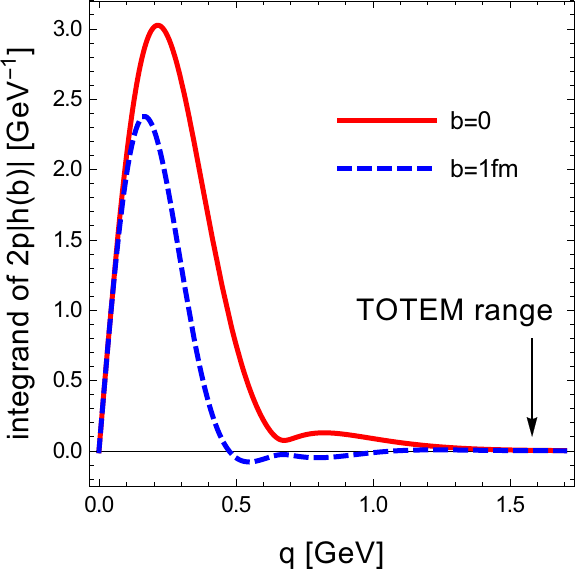}
\end{center}
\vspace{-5mm}
\caption{Absolute value of the integrand of the elastic amplitude from Eq.~(\ref{eq:invf}), $q J_0(bq) |f(s,-q^2)|/p$, plotted as a function of $q=\sqrt{-t}$ for two sample values of the impact parameter $b$.
The amplitude $f(s,-q^2)$ is taken from the parameterization~(\ref{eq:mBP2}) for  $\sqrt{s}=7$~TeV. The arrow indicates the upper range of the TOTEM~\cite{Antchev:2013gaa} 
data. \label{fig:integ}}
\end{figure} 

\begin{figure}
\begin{center}
\includegraphics[width=0.35\textwidth]{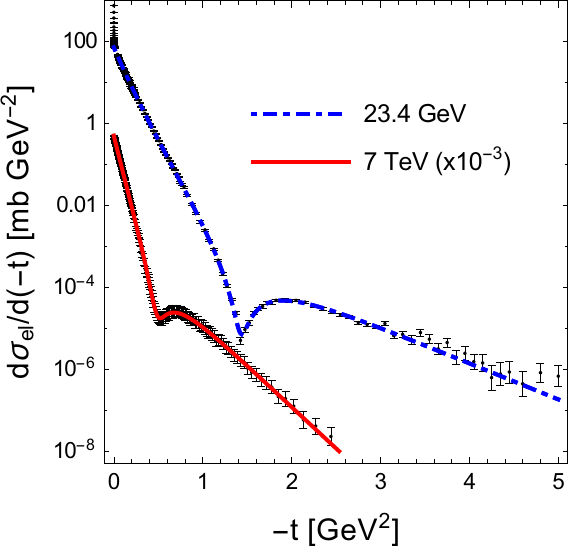}
\end{center}
\vspace{-5mm}
\caption{The differential elastic cross section in pp collisions from the MBP2 parametrization of Ref.~\cite{Fagundes:2013aja} (lines), 
compared to  the ISR~\cite{Amaldi:1979kd} data at $\sqrt{s}=23.4$~GeV and the
TOTEM~\cite{Antchev:2013gaa} data at $\sqrt{s}=7$~TeV.  \label{fig:247}}
\end{figure}

\begin{figure}
\begin{center}
\includegraphics[width=0.35\textwidth]{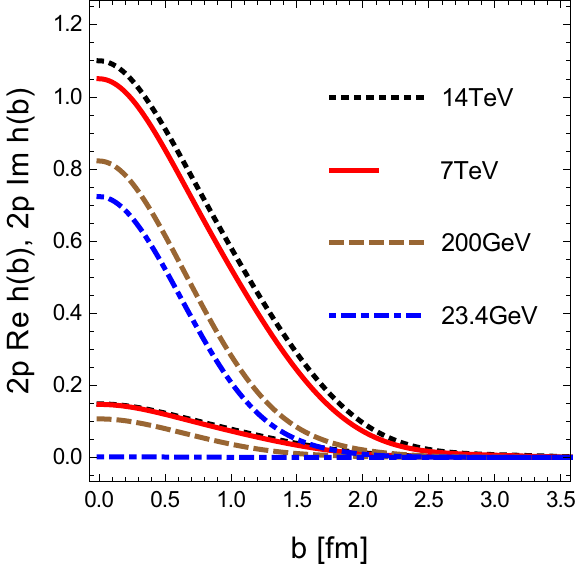}
\end{center}
\vspace{-5mm}
\caption{The imaginary (upper four curves) and real (lower four curves) parts of the eikonal
amplitude multiplied with twice the CM momentum, $2p h(b)$, plotted as functions of the impact parameter $b$. \label{fig:ph}}
\end{figure} 

\begin{figure*}
\begin{center}
\includegraphics[width=0.31\textwidth]{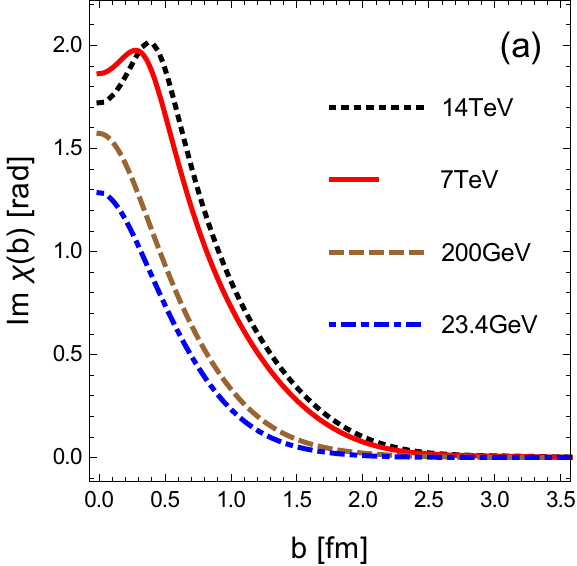}\hfill  \includegraphics[width=0.31\textwidth]{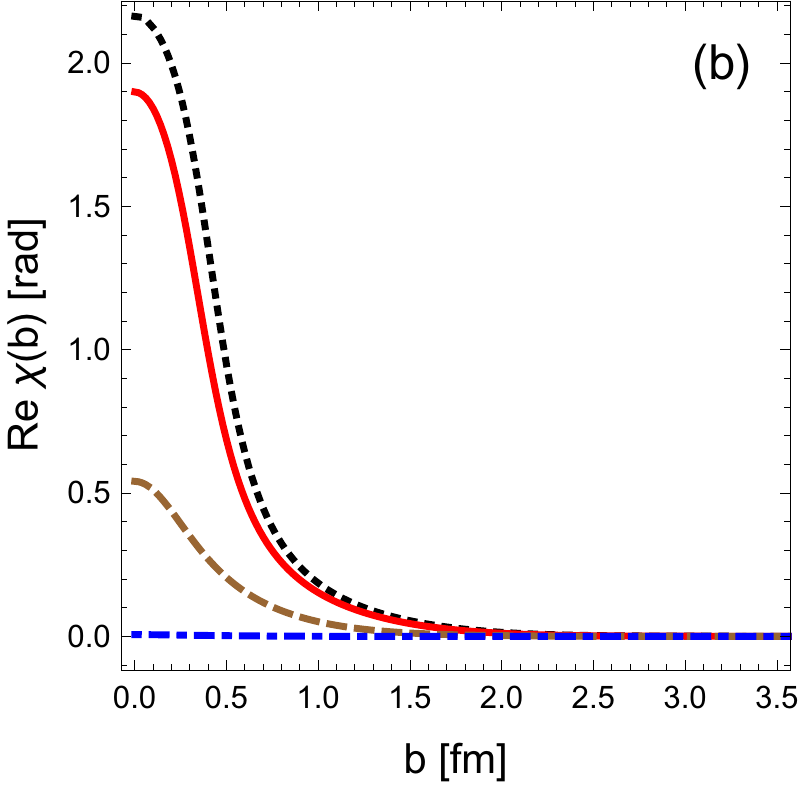} \hfill \includegraphics[width=0.31\textwidth]{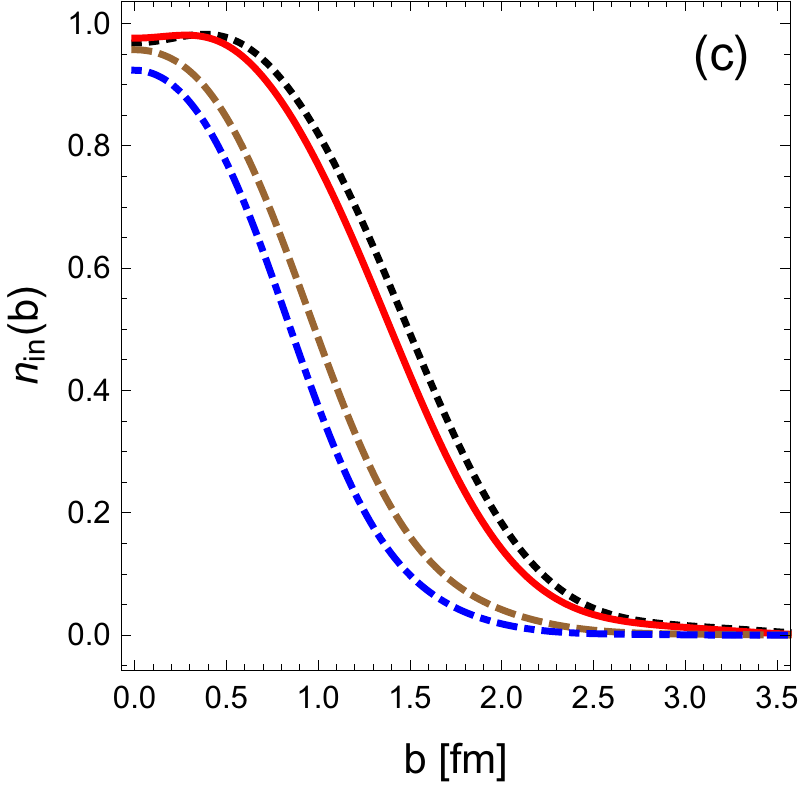}  \\
\vspace{3mm}
\includegraphics[width=0.31\textwidth]{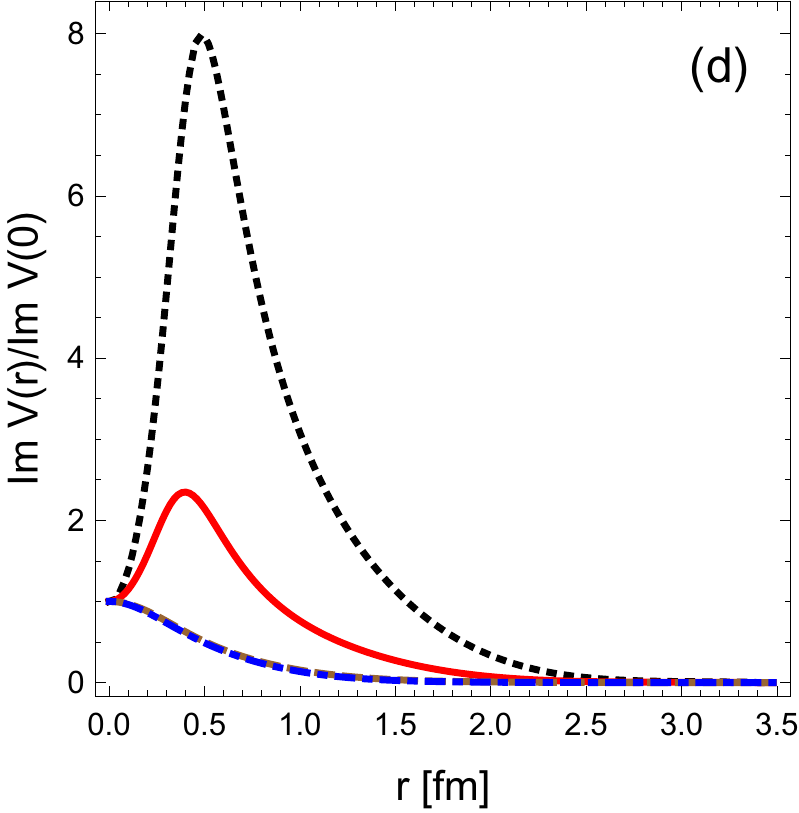} \hfill \includegraphics[width=0.31\textwidth]{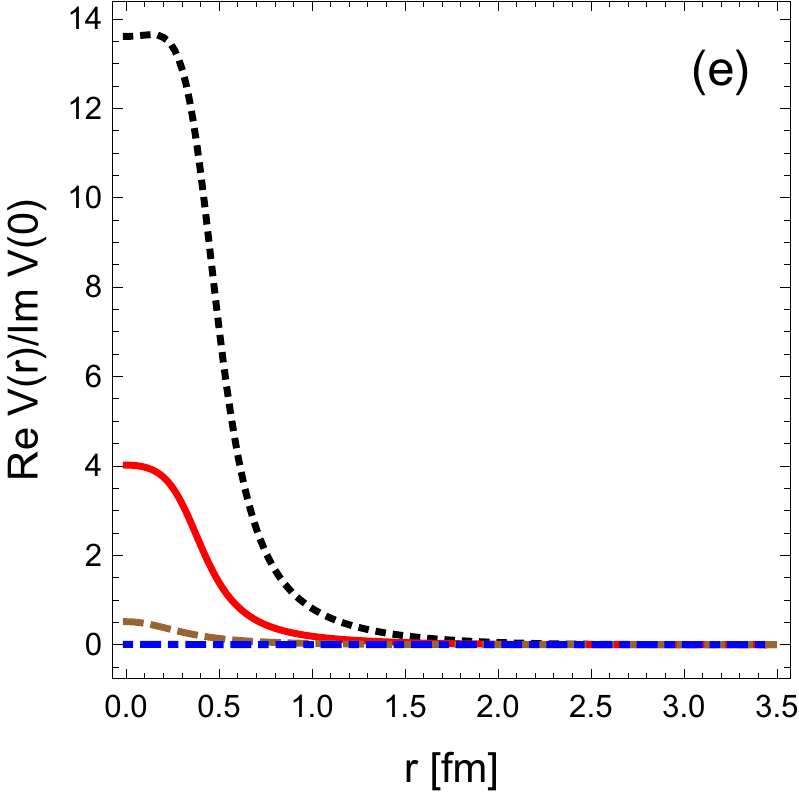}\hfill \includegraphics[width=0.31\textwidth]{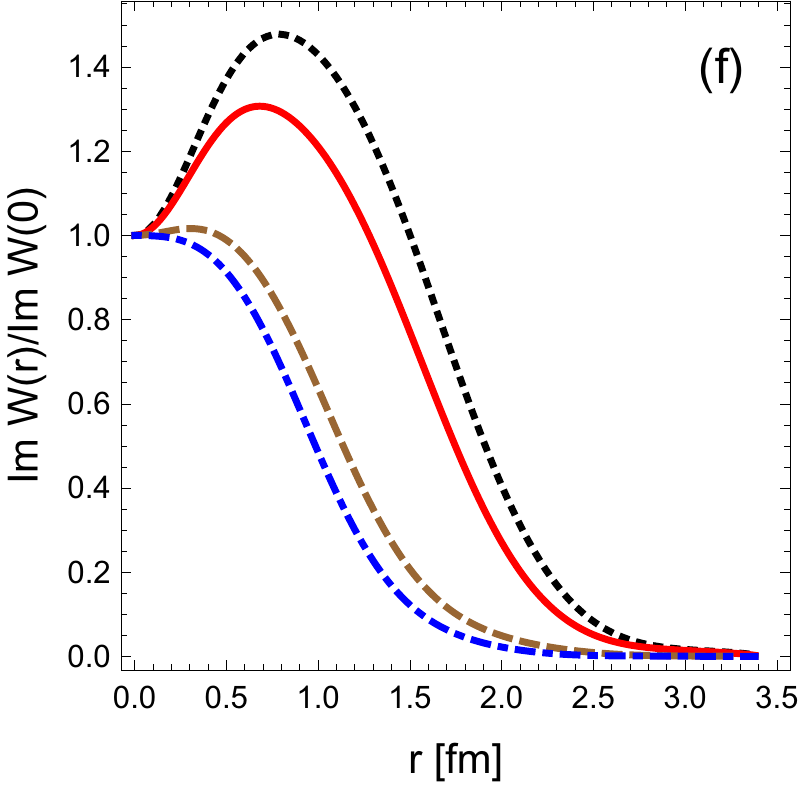}  \\
\end{center}
\vspace{-5mm}
\caption{
(a)~Imaginary part of the eikonal  phase $\chi(b)$, plotted as a function of $b$, for several collision energies.  
(b)~Same as (a) but for the real part of the eikonal phase. 
(c)~Same as (a) but for the  inelasticity profile $n_{\rm in}(b)$. 
(d)~Imaginary part of the optical potential $V(r)$ divided with ${\rm Im} V(0)$, plotted as a function of the radius $r$, for collision energies as in (a). 
(e)~Same as (d) but for the real part of the optical potential.
(e)~Same as (d) but for the imaginary part of the on-shell optical potential $W(r)$ divided with ${\rm Im} W(0)$. 
The plots in the lower row are obtained, correspondingly, from the plots in the upper row via the transformations of Eqs.~(\ref{eq:Uchi}) or (\ref{eq:Wn}).
\label{fig:all}}
\end{figure*}

We use the parametrization of the pp scattering data provided by
Fagundes {\it et al.}~\cite{Fagundes:2013aja} based in the
Barger-Phillips analysis~\cite{Phillips:1974vt} motivated by the Regge
asymptotics:
\begin{eqnarray}
{\cal A} (s,t) &\equiv& \frac{f(s,t)}{p} = \sum_{n} c_n(s) F_n(t) s^{\alpha_n (t)} \nonumber \\
&& \nonumber\\  && =\frac{i \sqrt{A} e^{\frac{B t}{2}}}{\left(1-\frac{t}{t_0}\right)^4}+i \sqrt{C} e^{\frac{D t}{2}+i \phi }, \label{eq:mBP2}
\end{eqnarray}
where the linear Regge trajectories $\alpha_n (t) = \alpha_n(0)+
\alpha_n'(0) t$ are assumed.  Specifically, we take the MBP2
parametrization of~\cite{Fagundes:2013aja}, with the $s$-dependent
parameters fitted separately to all known differential pp cross
sections for $\sqrt{s}= 23.4$, $30.5$, $44.6$, $52.8$, $62.0$, and
$7000~{\rm GeV}$ with a reasonable accuracy of $\chi^2/{\rm d.o.f}
\sim 1.2-4.7$.  A typical quality of the fit can be appreciated from
Fig.~\ref{fig:247}, where we show the comparison to the data at
two sample collision energies from ISR~\cite{Amaldi:1979kd} at
$\sqrt{s}=23.4$~GeV, and from the LHC (the TOTEM
Collaboration~\cite{Antchev:2013gaa}) at $\sqrt{s}=7$~TeV.  

\begin{table}[tb]
\caption{Scattering observables for different CM energy values, 
obtained from the MBP2 parameterization~\cite{Fagundes:2013aja} with
the inclusion of the $\rho$ parameter according to
Eq.~(\ref{eq:rhoind}), compared to experimental vales (lower rows). \medskip}
\begin{tabular}{cccccc}
 \hline
$\sqrt{s}$~[GeV] & $\sigma_{\rm el}$~[mb] & $\sigma_{\rm in}$~[mb] & $\sigma_{\rm T}$~[mb] & $B~[{\rm GeV}^{-2}]$ & $\rho$ \\ 
  \hline
 23.4 & 6.6 & 31.2 & 37.7 & 11.6 & 0.00 \\ 
\cite{Amaldi:1979kd} & 6.7(1) & 32.2(1) & 38.9(2) & 11.8(3) & 0.02(5) \\
  \hline
 200 & 10.0 & 40.9 & 50.9 & 14.4 & 0.13 \\ 
\cite{Aielli:2009ca,Bueltmann:2003gq} &  &  & 54(4) & 16.3(25) &  \\
  \hline
 7000 &  25.3 & 73.5 & 98.8 & 20.5 & 0.140 \\ 
\cite{Antchev:2013gaa} & 25.4(11) & 73.2(13) & 98.6(22)  & 19.9(3) & 0.145(100) \\
  \hline  
\end{tabular}
\label{tab:Fag-Mod}
\end{table}

To be consistent with the experimental values of the $\rho(s)$
parameter, where
\begin{eqnarray}
\rho(s) = \frac{{\rm Re}{\cal A} (s,0) }{{\rm Im}{\cal A} (s,0)},
\end{eqnarray}
we modify the amplitude of Eq.~(\ref{eq:mBP2}) be replacing it with 
\begin{eqnarray}
{\cal A} (s,t) \to \frac{i+\rho(s)}{\sqrt{1+\rho(s)^2}}|{\cal A} (s,t)|, \label{eq:rhoind}
\end{eqnarray}
which amounts to imposing a $t$-independent ratio of the real to
imaginary part of the amplitude. Other prescriptions have been
discussed in detail in Ref.~\cite{Antchev:2016vpy}.  We have checked
that our results are similar if we take the Bailly et
al.~\cite{Bailly:1987ki} parametrization
$\rho(s,t)=\rho_0(s)/(1-t/t_0(s))$, where $t_0(s)$ is the position of
the diffractive minimum.  Nevertheless, the results in the impact
parameter representation do depend to some extent on the form of
$\rho(s,t)$~\cite{Antchev:2016vpy} and the issue is intimately related
to the separation of the strong amplitude from the Coulomb part. As
these problems extend beyond the goals of this paper, we explore here
the simplest choice of Eq.~(\ref{eq:rhoind}).

Prescription~(\ref{eq:rhoind}) preserves the quality of the
fits of Fig.~\ref{fig:247}, and in addition the experimental values
for $\rho(s)$  are reproduced. For the explored below values of
$\sqrt{s}=23.4$~GeV, 200~GeV, 7~TeV, and 14~TeV we use,
correspondingly, $\rho=0$, 0.13, 0.14, and 0.135 (the last value is
obtained via extrapolation). For completeness, we provide
Table~\ref{tab:Fag-Mod} with the numerical results where predictions of  
Eq.~(\ref{eq:rhoind}) are compared to the available experimental
data.

Finally, we judge the accuracy of the eikonal approximation by
checking that the ratio $|h(b,p) / f_l(p)| \sim 1 $ to better than
$0.1\%$ when $b p=l+\frac12 $ and for $\sqrt{s} \ge 17~{\rm GeV} $\ and
$b \le 3~{\rm fm}$ for the MBP2 parameterization. The performance of
the approximation improves with increasing collision energy.

Before passing to the results, we also test whether the range of the data at
$\sqrt{s}=7$~TeV is sufficient to draw accurate conclusions for the
quantities in the impact-parameter representation. It is indeed the
case, as can be inferred from Fig.~\ref{fig:integ}.

\section{Results}

Our simple calculation consists of the following steps. First, with a
given parametrization for $f(s,t)$ we find $h(b)$ via a numerical
inverse Fourier-Bessel transform in Eq.~(\ref{eq:PWA}). Then from
Eq.~(\ref{eq:eik}) we obtain the eikonal phase and the quantities from
Eq.~(\ref{eq:st}-\ref{eq:prof}), whereas the optical potentials follow
from Eqs.~(\ref{eq:Uchi},\ref{eq:Wn}).  The relevant quantities are
displayed in Figs.~\ref{fig:ph} and \ref{fig:all}. A few characteristic features
should be stressed.

First, we note from Fig.~\ref{fig:ph} that with an increasing
collision energy from ISR via RHIC to the LHC, the real part of the
eikonal scattering amplitude ${\rm Re} h(b)$, while remaining small,
increases (in our model, simply, ${\rm Re}\, h(b)=\rho {\rm Im}\, h(b)$). At the
LHC energies, it reaches $\sim 15\%$ percent of the dominant imaginary
part.

The eikonal phase is presented in Figs.~\ref{fig:all}~(a, b). We can
see that its imaginary part develops a dip at the origin at the LHC
energies. Moreover, it achieves a very sizable positive real part, of the
size of the imaginary part at the LHC.

The inelasticity profile $n_{\rm in}(b)$, Fig.~\ref{fig:all}~(c), flattens
near the origin as the collision energy is being increased, and for
the LHC develops a shallow minimum at $b=0$, whereas the maximum shifts to
$b>0$. Note that by construction and in accordance to unitarity
$n_{\rm in}(b) \le 1$. The dip at $b=0$ is a symptom of the 2D hollowness effect,
discussed in a greater detain in the next section.

Finally, we observe dips in the imaginary parts of both the optical
potential $V(r)$, Fig.~\ref{fig:all}~(d), and the on-shell optical
potential $W(r)$, Fig.~\ref{fig:all}~(f), appearing prominently with
an increasings $s$ and displaying the hollowness effects in 3D.

\section{The nature of the hollow}

\begin{figure}[ttt]
\begin{center}
\includegraphics[width=0.35\textwidth]{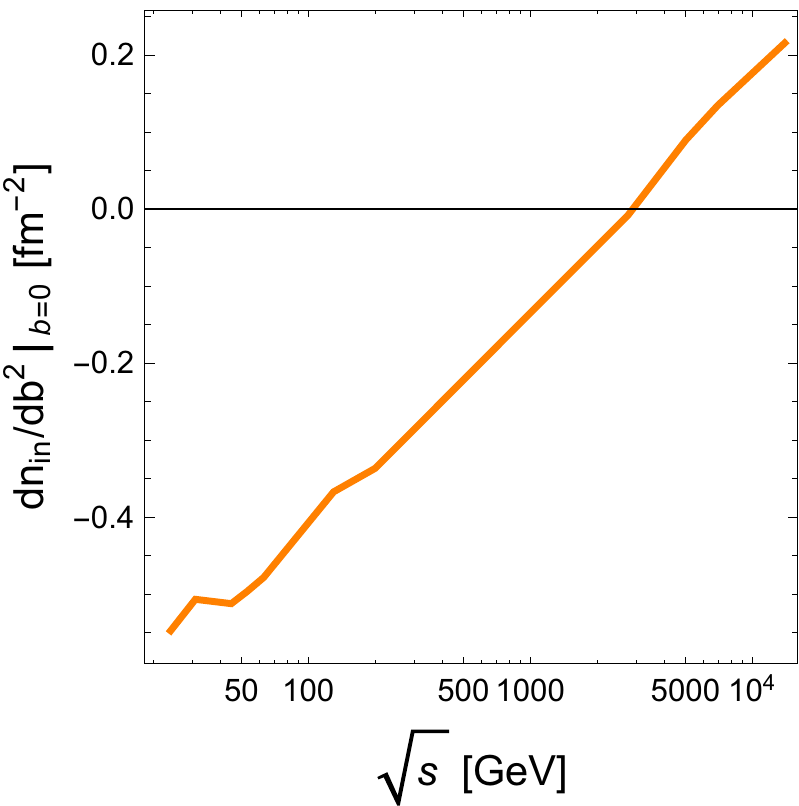}
\end{center}
\vspace{-5mm}
\caption{The curvature of the inelasticity profile $n_{\rm in}(b)$ at $b=0$ plotted as 
a function of the collision energy. The minimum emerges at $\sqrt{s} \simeq 3$~TeV. \label{fig:slope}}
\end{figure}

As the pp collision energy increases, the total inelastic cross
section $\sigma_{\rm in}(s)$ grows. Moreover, as shown in the previous
section, the inelasticity profile in the impact parameter flattens at
the origin, or even develops a shallow minimum at sufficiently large
$s$, as follows from Fig.~\ref{fig:slope}. By simple geometric
arguments, this flattening must correspond to a 3D hollow in the radial
density of inelasticity, here interpreted as the on-shell optical
potential ${\rm Im} W(r)$, cf. Eq.~(\ref{eq:nin1}). In fact, for the
collision energies above the lowest ISR case of $\sqrt{s}=24.3$~GeV
the function ${\rm Im} W(r)$ exhibits a depletion at the origin -- the
hollow.  

As depicted in the introductory Fig.~\ref{fig:3D2D},
the ``hollowness'' effect is more pronounced when interpreted in 3D,
i.e., via ${\rm Im}W (r)$, than in its 2D projection, namely $n_{\rm in}(b)$
(cf. Eq.~(\ref{eq:nin1})), since a 3D function is integrated over the
longitudinal direction, which effectively covers up the hole. 

Folding ideas have been implemented in microscopic models based on
intuitive geometric
interpretation~\cite{Chou:1968bc,Chou:1968bg,Cheng:1987ga,Bourrely:1978da,Block:2015sea}. Interestingly,
the 3D hollowness effect cannot be reproduced by naive folding of
inelasticities of uncorrelated partonic constituents. If $\Psi_A
(x_1,x_2,x_3, \dots )$ and $\Psi_B (x'_1,x'_2,x'_3, \dots )$ are the
corresponding partonic wave functions of hadrons A and B, the single
parton distributions are given by
\begin{eqnarray}
\rho_A (\vec x_1 ) &=& \int d^3 x_2 d^3 x_3 \dots | \Psi_A (x_1,x_2,x_3, \dots ) |^2, \nonumber \\  
\rho_B (\vec x_1') &=& \int d^3 x'_2 d^3 x'_3 \dots | \Psi_B (x'_1,x'_2,x'_3, \dots ) |^2. 
\end{eqnarray}
In a folding model, antisymmetry of the wave functions is neglected and
the absorptive part of the potential, ${\rm Im} \, W(r)$, is
proportional to the overlap integral
\begin{eqnarray}
&&\langle \Psi_A \Psi_B | \sum_{i \in A,i' \in B} w ( \vec x_i - \vec x_{i'} - \vec r)  |  \Psi_A \Psi_B \rangle \nonumber \\ &=& \int d^3 y \rho_A ( \vec x + \vec r/2) w(\vec x' - \vec x ) \rho_B (\vec x-\vec r/2) \, . 
\end{eqnarray}
where $w(\vec x-\vec x')$ denotes the interaction among constituents
belonging to different hadrons (we omit further possible
indices). For identical hadrons, A=B, and at small $r$ we get
\begin{eqnarray}
{\rm Im}\,W(r) & \propto & \int d^3 x d^3 x' 
\rho( \vec x' + \vec r/2) w(\vec x' - \vec x ) \rho(\vec x-\vec r/2) \nonumber \\ &=& \int d^3 y d^3 y' \rho(\vec x) \rho(\vec x') w(\vec x' - \vec x )  \nonumber \\ 
&-& \frac12  \int d^3 x d^3 x' [\vec r \cdot \nabla \rho(\vec y)]
w(\vec x-\vec x') [\vec r \cdot \nabla' \rho(\vec x')] \nonumber \\
&+& \dots ,
\end{eqnarray}
For a positive $w(\vec x-\vec x')$ both integrals are necessarily
positive as can be seen by going to the Fourier space. This proves that if
${\rm Im}W(r)$ stems from a folding of densities with $w(\vec x-\vec
x')>0$, then it necessarily has a local maximum at $r=0$, in contrast
to the phenomenological hollowness result. Folding models usually take
$w(\vec x-\vec x') \propto \delta (\vec x' -\vec
x)$~\cite{Chou:1968bc,Chou:1968bg,Cheng:1987ga,Bourrely:1978da,Block:2015sea}.

Note that the above conclusion holds for any wave functions
$\Psi_{A,B}$, correlated or not. In particular, one may think of
modeling collisions of objects empty in the middle (for instance,
protons made as triangles of three constituents).  If inelasticity
were to be obtained via above density folding, even in this case the
absorption would be strongest for head-on collisions.

Likewise, the 2D hollowness cannot be obtained by folding structures
in the impact parameter space, as for instance used in the models of
Ref.~\cite{Chou:1968bc,Chou:1968bg,Cheng:1987ga,Bourrely:1978da,Block:2015sea}.

In our model one may give a simple criterion for $n_{\rm in}$ to develop
a dip at the origin. Introducing the short-hand notation 
$k(b)=2 p {\rm Im}\, h (b)$ we have immediately from
Eqs.~(\ref{eq:prof}) and (\ref{eq:rhoind}) the equality 
\begin{eqnarray}
n_{\rm in}(b)=k(b)-(1+\rho^2) k(b)^2/2. 
\end{eqnarray}
Differentiating with respect to $b^2$ one immediately finds that
$d^2n_{\rm in}(b)/db^2$ at the origin is negative when
\begin{eqnarray}
2 p {\rm Im} h (0)>\frac{1}{1+\rho^2}\sim 1, \label{eq:crit}
\end{eqnarray}
where the departure from 1 is at a level of $2\%$ at the LHC and
smaller at lower collision energies.

One can make the following direct connection to the eikonal
phase. From Eq.~(\ref{eq:eik}) we get immediately
\begin{eqnarray}
2p{\rm Im}\, h(b)= 1-\cos \left ( {\rm Re}\chi(b) \right ) e^{-{\rm Im}\chi(b)}, \label{eq:eik2}
\end{eqnarray}
hence $2p{\rm Im}h(0)>1$ (thus satisfying
criterion~(\ref{eq:crit})) when ${\rm Re}\chi(0)$ increases above
$\pi/2$, whence 
\begin{eqnarray}
\cos({\rm Re} \chi(b))<0. \label{eq:cos}
\end{eqnarray}
This is indeed the case in Fig.~\ref{fig:all}(b).  

Recall that in the Glauber model~\cite{glauber1959high} of scattering of
composite objects, the eikonal amplitudes of individual scatterers are
additive, composing the full eikonal amplitude $\chi(b)$. 
Thus, in this quantum-mechanical framework, the monotonic change of $\chi(b)$
with the collision energy may be caused by the corresponding change of
the eikonal amplitudes of the scatterers, or the increase of the
number of scatterers (as expected from the growing number of gluons at
increasing energies), or both.  Thus a quantum nature of the
scattering process is the alleged key to the understanding of the
hollowness effect. 

Finally, we show that the 2D flatness of the inelasticity profile
$n_{\rm in}(b)$ at the origin implies the 3D hollowness in ${\rm Im}
W(r)$.  If $n_{\rm in}(b) $ is constant for $0 < b \le b_0$, then
lower range of the integration in Eq.~(\ref{eq:Wn}) starts from $b_0$
and the integral has no singularity for $r < b_0$. Direct inspection
shows that the magnitude of ${\rm Im} W (r)$ grows with $r$, which
corresponds to 3D hollowness.

\section{The hollow and the edge}

\begin{figure}[tb]
\begin{center}
\includegraphics[width=0.35\textwidth]{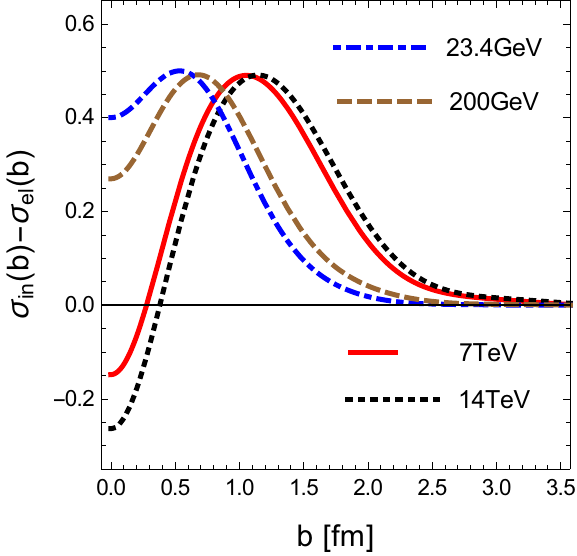}
\end{center}
\vspace{-5mm}
\caption{The edge function $\sigma_{T}(b) - 2 \sigma_{\rm el} (b) \equiv 
\sigma_{\rm in}(b)-\sigma_{\rm el} (b)$ as a function of the impact parameter
for different energies.  \label{fig:edge}}
\end{figure}

The {\em edge} function, based on defining $\eta(b)= e^{-{\rm
Im} \chi(b)}$ and analyzing the combination $\eta (b) (1-\eta (b))$,
has been considered in Refs.~\cite{Block:2014lna,Block:2015sea} (see,
e.g., Ref.~\cite{Rosner:2014nka} for an interpretation in terms of
string breaking).  In the limit of a purely imaginary amplitudes the
edge function can be interpreted as a combination of the unintegrated
cross sections $\sigma_{T}(b) - 2 \sigma_{\rm el}
(b) \equiv \sigma_{\rm in}(b)-\sigma_{\rm el} (b)$ (we use $\sigma_{\rm
in}(b) \equiv n_{\rm in}(b)$). In the general case, with the real part
of the eikonal phase present, it reads
\begin{eqnarray}
\sigma_{\rm in}(b)-\sigma_{\rm el} (b) =2 e^{-{\rm Im}\chi(b)} \left[ \cos({\rm Re} \chi(b)) -e^{-{\rm Im}\chi(b)} \right]. \nonumber \\
\end{eqnarray}
We show in Fig.~\ref{fig:edge} the edge functions at various collision
energies resulting from our analysis. As we can see, the edge function
becomes negative at the LHC energies for $b \lesssim 0.5 {\rm fm}$ due
to the same quantum mechanical effect as described in the previous
section, leading to condition~(\ref{eq:cos}). Note that in
Ref.~\cite{Block:2015sea} there is no region in $b$ with a negative
contribution in the edge function because of the folding nature of the
underlying model. Instead, one observes a 2D flatness, which complies,
according to our analysis, to a 3D hollowness.

Therefore the fact that $\sigma_{\rm in} (b) < \sigma_{\rm el} (b) $
at low $b$ at the LHC energies provides an equivalent manifestation of the
hollowness effect. In other words, at low impact parameters there the
unintegrated inelastic cross section is smaller from its elastic
counterpart.

\section{Conclusions}

Over the past years many analyses have tempted to regard the largest
available energy as close enough to the asymptotia holy grail, but so
far this expectation has been recurrently frustrated. The
new LHC data on pp scattering may suggest a change in a basic paradigm 
of high energy collisions, where the head-on ($b=0$) collisions are 
expected to create ``more damage'' to the system compared to collisions 
at $b>0$. 

We have shown that a working parametrization of the pp scattering data
at the LHC energies indicates a hollow in the inelasticity profile
$n_{\rm in}(b)$, i.e. a dip at the origin, confirming the original
ideas by Dremin~\cite{Dremin:2014eva,Dremin:2014spa}. In other words,
there is less absorption for head-on collisions ($b=0$) than at a
non-zero $b$.  The shallow dip found from parameterizing the present
data at $\sqrt{s_{NN}}=7$~TeV is subject to experimental uncertainties
and, to some extent, on assumptions concerning the ratio or the real
to imaginary parts of the scattering amplitudes as functions of the
momentum transfer. Nevertheless, its emergence, if confirmed by future
data analyses at yet higher collision energies, has far-reaching
theoretical consequences. We have shown with a simple geometric
argument that in approaches which model the inelasticity profile by
folding partonic densities of the colliding protons, the 2D hollowness
is impossible.

We have used techniques of the inverse scattering in the eikonal
approximation to show that the optical potential and the on-shell
optical potential display the hollowness effect in 3D much more
vividly than the 2D inelasticity profile in the impact parameter
space.  The 2D hollowness will presumably be more pronounced at higher
collision energies, but in 3D it sets in at much lower energies than
the LHC.  Our approach gives a spatial insight into the three
dimensional geometric structure of the inelasticity region.  The found
hollowness in 3D, which is a robust effect, contradicts an
interpretation of the absorptive part of the on-shell optical
potential via naive folding of partonic densities.

A final confirmation of the 2D hollowness requires more detailed
studies both on the experimental as well on the theoretical side. In
contrast, the presence of the 3D hollowness can be established from a
flat behavior of the inelasticity profile in the small $b$ region,
which is estimated to happen at the LHC energies. Our inverse
scattering approach yields, however, that the 3D hollowness transition
takes place already within the ISR energy range.

We have argued that the hollowness effect has a quantum nature which
may be linked to the interference if the scattering of constituents in
the Glauber framework.  In 2D in the eikonal approximation, hollowness
occurs when the real part of the eikonal scattering phase goes above
$\pi/2$.

Furthermore, we have also pointed out that the 2D hollowness is
intimately linked to the edginess; moreover, with the used
parametrization of the data, the inelastic profile at low impact
parameters is smaller than its elastic counterpart, causing the edge
function to become negative.

\bigskip

We thank Alba Soto Ontoso and Javier Albacete for discussions.  This
work was supported by the Spanish Mineco (Grant FIS2014-59386-P), the
Junta de Andaluc\'{\i}a (grant FQM225-05), and by the Polish National
Science Center grants DEC-2015/19/B/ST2/00937 and
DEC-2012/06/A/ST2/00390. ERA acknowledges grant of he Polish National Science
Center 2015/17/B/ST2/01838.


%

\end{document}